\documentclass[prl,twocolumn,showpacs,groupedaddress]{revtex4}
\usepackage{epsfig}
\begin{document}
\title{Spin-flip Effects in the Mesoscopic Spin-Interferometer.} 
\author{A.Aldea$^{*,\dagger}$, M.Tolea$^*$ and J.Zittartz$^\dagger$}
\affiliation{$^*$ National Institute of Materials Physics, POBox MG7,
Bucharest-Magurele, Romania \\
$^\dagger$ Institut f\"ur Theoretische Physik, Universit\"at zu K\"oln,
D-50923 K\"oln, Germany}

\begin{abstract}
We investigate the properties of the electron spin-transmission
through an Aharonov-Bohm interferometer with  an embedded two-
dimensional multilevel quantum dot containing magnetic impurities. 
A suitable formalism is developed.
The amplitude and the phase of the flip- and nonflip-transmittance are
calculated numerically as function of the magnetic field and 
the gate potential applied on the dot.
The effects induced  by the exchange interaction to
spin-dependent magnetoconductance fluctuations and 
transmittance phase are shown. 
\end{abstract}
\maketitle
\section {I. Introduction}
The spin interferometry in mesoscopic systems is expected to give
new insights in the field of nano-physics.
We approach this  problem in a
ring-dot geometry with the aim to  identify 
new properties of the Aharonov-Bohm oscillations due to the
presence of  magnetic impurities within a multiple level dot;
we show also the ensemble of parameters that determine the problem.
\par
The Aharonov-Bohm interferometer with embedded quantum dot
has been used to study  the phase of the electron transmission
as a voltage applied on the dot is varied \cite{Ji,Weiden}, 
however  no attention was paid till now to the spin transmittance 
or even to magnetoconductance fluctuations in such tunable  systems, 
where the finite size of the dot plays an effective role.
The set-up is sketched in Fig.1;  the dot  contains magnetic impurities,
which implies an exchange interaction between the spin of the 
free incident electron and the localized spins in the dot.
In the real systems, the dot has a finite size so that the
electron crosses the dot along different paths, with implications for
the resulting interference pattern.
This is why the usual model which assimilates the dot with a one-site
impurity (with at most two-orbitals) \cite{IZ} cannot capture
all the interference  effects in actual ring-dot systems.
The price to be paid for considering a  realistic geometry for the dot
is to make a single electron approach; we mention in this respect
the recent paper by Nakanishi et al \cite{Naka} who uses a continuous
model for the study of the Fano effects.
\begin{figure}
\vspace*{-13.5cm}
\hspace*{2.5cm}
\vspace*{0.2cm}
\epsfxsize 6 cm
\epsffile{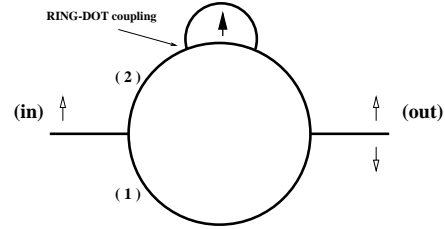}
\vspace*{5.0cm}
\caption
{The sketch of the spin interferometer; the incident electron is 
spin-polarized; the dot contains a localized magnetic impurity which 
couples with the spin of the free electron. 
A magnetic field pierces perpendicularly the ring and the dot. }
\end{figure}

The spin-dependent transmittance amplitude can be written as a superposition
of all possible paths:
\begin{equation}
t_{\sigma,\sigma'}=t^{1} \delta_{\sigma,\sigma'}+\sum_{k\in 2}
t^k_{\sigma,\sigma'}~e^{i(\Phi+\Delta\phi^k_{\sigma,\sigma'})},
\end{equation}
where $t^1$ is the amplitude of the transmittance through the lower
arm which conserves the spin, $\Phi$ is the magnetic flux that 
pierces the ring and $\Delta \phi$ is the
supplementary contribution  of the dot, where spin-flip processes may  occur.


The aspect of the output depends on the magnetic field that pierces
both the ring and  the dot and on the gate potential applied on the dot;
at the same time, it is  strongly controlled by the strength 
of the ring-dot coupling ($\tau$) and  the energy of the incident
electron($E_F$).

As explained throughout the paper, the range of parameters is large 
and different regimes can  be studied.
For instance, in the case of small $\tau$, one may consider that the ring-dot
coupling only produces broadening and  slight shifts in the 
spectrum of the individual  dot.
In the opposite case, this picture is no longer valid  and 
the ring-dot system has to be regarded as a unique coherent system, 
meaning that $\tau$ must be considered in all powers of the 
perturbation series. 
The hybridization depends on the electric potential $V_g$ 
since the applied potential shifts the dot levels.
However, for a two-dimensional quantum dot, the same can be 
achieved by the variation of the magnetic flux,
as put into evidence by a recent experiment carried out in the
same ring-dot geometry \cite{Aikawa}. 
We shall focus our study on the spin dependent magnetoconductance as 
function of flux at fixed gate potential.

Since several energy scales  come into the problem, 
one has to establish from the very beginning the assumptions of the 
approach.
We consider that the main processes which are to be taken into 
account are the orbital motion in magnetic field along the ring 
and inside the dot, and the  exchange interaction $J$.
We assume that the Zeeman effect is less effective than the exchange, i.e.
$g\mu_B~B /J <1$~~\cite{comm} ; 
the spin-orbit interaction is neglected, too.
We use a tight-binding (TB) description that proved to be efficient in
describing interference effects for differently shaped systems and in
the presence of disorder and interactions ~\cite{AGC,YB}.
The magnetic field (in the Landau gauge) appears as a Peierls phase $\Phi_{ij}$
in the hopping integral, and only the next-neighbor hopping is considered.
The dot is an  island of $3\times 5$ sites
attached on the external side of the ring, with the magnetic impurity 
placed in the middle. The dot area represents $15\%$ from the area
of the whole device.
The Hamiltonian of the ring-dot system in perpendicular magnetic field reads:
\begin{eqnarray}
H= \sum_{\langle i,j\rangle,\sigma}w_{ij,\sigma}~ e^{i2\pi
\Phi_{ij}}c_{i,\sigma}^\dagger c_{j,\sigma}  +
\sum_{\sigma,i\in QD}V_g c_{i,\sigma}^{\dagger} c_{i,\sigma}-\nonumber\\ 
-J~ (c^\dagger_{n\uparrow} c_{n\uparrow}-
c^\dagger_{n\downarrow} c_{n\downarrow})S^z_n -
J~(c^\dagger_{n\downarrow} c_{n\uparrow}S^+_n +H.c.), \nonumber \\
~~~~~~~~~ n\in QD.
\end{eqnarray}
where ~$c_i^\dagger~ (c_i)$ are creation (annihilation) operators in 
localized states indexed by $i$~;
 the index $n$ is devoted to the site of the dot where the magnetic
impurity -called also the 'flipper'- is placed.
The first term describes both the 1D ring and the dot ($i,j=1..N$), the second
term allows for the gate potential on the dot and the last terms
represent the local spin-spin interaction.
The hopping integral will be taken as  energy unit, i.e. 
$w_{ij,\sigma}=1$ for any pair (i,j) excepting at the contacts between 
the dot and the ring,  where $w=\tau \in[0,1]$.

The outline of this paper is as follows. In section II  
we adapt the formalism used previously for the spinless problem \cite{MAMN}
to the spin transport through meso systems.
This yields a fast way to obtain numerical results in the presence of
spin scattering based on the calculation of the resolvent in a 
reduced Hilbert space and Landauer-B\"uttiker formalism. 
The existence of different regimes is shown.
Section III introduces the singlet and triplet operators and expresses 
the flip and non-flip processes in terms of two-time Green (singlet \&triplet) 
functions with their corresponding  equations of motion. 
We also show how the single electron scattering problem is recovered 
from the general many-body scheme.
An application is done for an analytically soluble model in 
sec IV, with the aim to produce hints for the  ring+2Ddot
interferometer. This complex system is studied numerically in sec V, where
spectral and transport properties are shown, indicating the role
of different parameters like the Fermi energy and the  coupling constant
$\tau$. The main conclusions are presented in the last section.
\section{II. Spin tunneling in the resolvent representation and discussion of
different regimes}
In what follows we shall reduce the many-body problem to the
physical situation when a "test" electron carrying the spin $\sigma$
passes through the interferometer and interacts via the exchange $J$
with the localized spin $S$. So we have to project the Hamiltonain on the
product Hilbert space describing a single free spin and a single localized spin:
$\mathcal H^{\sigma}_{2N}~\otimes~\mathcal H^S_{2S+1}$. This space is spanned
by the basis ${\{|i,\sigma>\}}\otimes{\{\chi_S\}}$. For this reason, in the
present section we shall use the bra/ket representation instead of the
creation/annihilation operators.

For the calculation of the transmittance (in the Landauer-B\"uttiker formalism)
additional terms are needed  in order to describe 
the external leads and their contacts to the ring-dot system.
It is however sufficient to use an effective  Hamiltonian depending only
on the  degrees of freedom of the ring-dot, but containing 
the whole information about the leads and lead-ring coupling in a 
non-Hermitian term \cite{MAMN}:
\begin{equation}
H^{eff}= H + \tau_0^2 \sum_{\alpha,\sigma} e^{-iq}~ |\alpha\sigma> 
<\alpha\sigma| , 
\end{equation}
where $q$ is defined by $E_F=2 cos( q)$ and \{$\alpha$\} denotes the sites where
the leads are connected to the system. $\tau_{0}$ is the parameter
describing the lead-ring coupling and it will be taken $\tau_{0}=1$.
Taking into account that the lead-ring coupling conserves the spin, only the
diagonal elements are modified by the coupling $\tau_0$, and one can define
the non-flip effective Hamiltonian :
\begin{eqnarray} 
H_{\uparrow\uparrow}^{eff}=
\sum_{\langle i,j\rangle}w_{ij,\uparrow}~ e^{i2\pi
\Phi_{ij}}|i\uparrow><j\uparrow|+\sum_{i\in QD}V_g |i\uparrow>
<i\uparrow|\nonumber\\
-JS_n^z~ |n\uparrow><n \uparrow|+\tau_{0}^2\sum\limits_\alpha e^{-iq}~
|\alpha\uparrow><\alpha\uparrow|~, ~~~~~~
\end{eqnarray}
while the spin-flip Hamiltonian is simply:
\begin{equation}
H_{\uparrow\downarrow} ^{eff}=H_{\uparrow\downarrow}=
-JS_n^-|n\uparrow><n\downarrow|.
\end{equation}
In order to calculate the spin transmittance through the complex system 
one needs to know the retarded resolvent Green function:
$$G^+(E)=(E-H^{eff}+i0)^{-1}.$$
Then, in order to 
express the different flip and non-flip 
tunneling processes  we define the 2x2 matrix Green function
$$ G_{\sigma,\sigma'}(z)=:<\sigma| G(z)|\sigma'>$$ for which Dyson equations 
can be written immediately and the expressions for the nonflip 
$G_{\uparrow\uparrow}(z)$ and flip $G_{\uparrow\downarrow}(z)$ read :
\begin{eqnarray}
G_{\uparrow\uparrow}(z)=\Big(z-H_{\uparrow\uparrow}^{eff}-
H_{\uparrow\downarrow}(z-H_{\downarrow\downarrow}^{eff})^{-1}
H_{\downarrow\uparrow}\Big)^{-1}  \nonumber \\
\hskip -0cm G_{\uparrow\downarrow}(z)=G_{\uparrow\uparrow}(z)
H_{\uparrow\downarrow}^{eff}
(z-H_{\downarrow\downarrow}^{eff})^{-1}.~~~~~~~~~~~~~~~~
\end{eqnarray}
Using (5) these equations can be further expressed as:
\begin{eqnarray}
G_{\uparrow\uparrow}(z)=\Big(z-H_{\uparrow\uparrow}^{eff}-J{^2}S_n^-
|n\uparrow> \nonumber ~~~~~~~~~~~~~~~~ \\
< n\downarrow|{1\over z-H_{\downarrow\downarrow}^{eff} }
|n\downarrow><n\uparrow|S_n^+ \Big)^{-1} ~,\nonumber\\
G_{\uparrow\downarrow}(z)=-J~G_{\uparrow\uparrow}(z)S_n^-
|n\uparrow><n\downarrow|{1\over z-H_{\downarrow\downarrow}^{eff}}~ .
\end{eqnarray}

In order to proceed we shall consider $S_n=1/2$.
Since the exchange interaction conserves the total spin and its projection
on the z-axis, there are three different tunneling processes between the
following $in$ and $out$ states:
\begin{eqnarray}
a)~~ |\alpha\uparrow\Uparrow> ~\longrightarrow~ |\alpha'\uparrow \Uparrow>
~~(non-flip~process)\nonumber\\
b)~~ |\alpha\uparrow \Downarrow>~ \longrightarrow~ |\alpha'\uparrow \Downarrow>
~~(non-flip~process)\\
c)~~ |\alpha\uparrow \Downarrow>~ \longrightarrow~ 
|\alpha'\downarrow \Uparrow> ~~(flip~process)\nonumber~~~~~~~~~
\end{eqnarray}
The first two processes contribute to the non-flip transmittance 
while the third process to the flip transmittance.
We remind that the pair of sites $(\alpha, \alpha')$  represent the points
where the ring is connected to the terminals and that the incoming spin
is supposed to be polarized in the $up(\uparrow)$-state.
The notation $|\Uparrow>,|\Downarrow>$ stands for the eigenvectors of the
impurity spin operator $S_n^z$.

Finally, the spin transmittance is obtained by calculating
the matrix elements of  $G_{\sigma,\sigma'}$
in (7) between these  {\it in} and {\it out} states, and using them in
the general expression of the Landauer-B\"uttiker formula: 
\begin{equation}
T_{\sigma,\sigma'S,S'}(E,\Phi)=
4 \tau_0^4 {sin^2}q ~|<\alpha,S|G^+_{\sigma \sigma'}(E)|
\alpha',S'>|^2
\end{equation}
As an explicit example of how the matrix elements look like, we give :
\begin{eqnarray}
<\alpha\uparrow\Uparrow|~G_{\sigma\sigma}(z)|\alpha'\uparrow\Uparrow>=~~~~
~~~~~~~~~~~~~~~~~~~~~~~~~~~~~~~~\nonumber \\
<\alpha\uparrow|~\Big(z-\sum_{\langle i,j\rangle}w_{ij,\uparrow}~ e^{i2\pi
\Phi_{ij}}|i\uparrow><j\uparrow| - ~~~~~~~~~~~~~ \nonumber \\
{\tau_0}^2~ e^{ik}\sum_{\alpha_1}| \alpha_1\uparrow><\alpha_1\uparrow|+
{J\over2}~|n\uparrow><n\uparrow\Big)^{-1}|\alpha'\uparrow>.\nonumber\\
\end{eqnarray}
The  value of this Green function element can be calculated by the 
direct numerical inversion of the matrix which is now completely known.

The Aharonov-Bohm interference is impossible when the dot does not transmit,
since the circulation of the vector potential is to be considered  along a 
closed contour. 
The transmittance of the dot can be controlled  by changing the gate 
potential applied on the dot.
Therefore, the transmittance spectrum in the plane of the variables 
$V_g$ (gate potential) and $\Phi$ (magnetic flux through the ring) 
has to be analyzed  first.

\begin{figure}
\vspace*{-14.2cm}
\hspace{-4.0cm}
\epsfxsize 9.0 cm
\epsffile{s_fig2a.ps}
\vspace{9.2cm}
\end{figure}
\begin{figure}
\vspace*{-18.0cm}
\hspace{-4.0cm}
\epsfxsize 9.0 cm
\epsffile{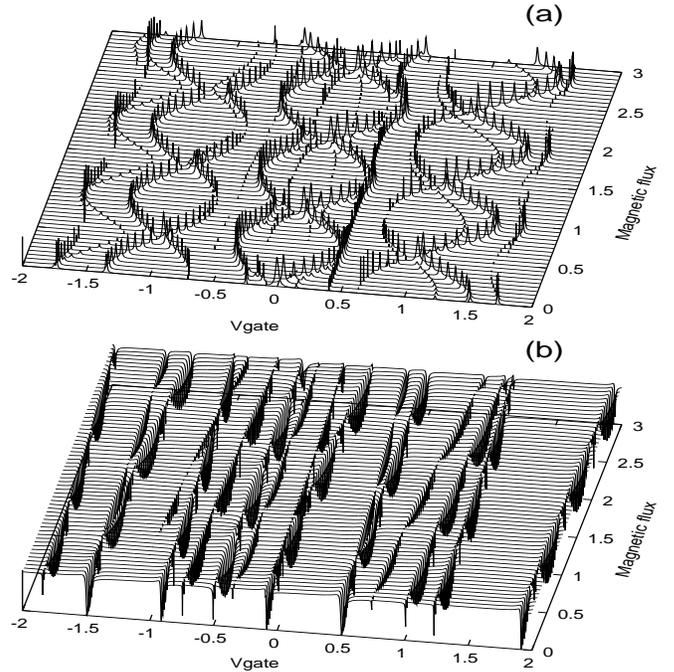}
\vspace*{5.2cm}
\caption
{The map of the nonflip-transmittance in the plane ($V_{gate}$, Magnetic flux)
for two values of the Fermi energy: (a) $E_F=0$  and (b) $E_F=0.5$.
The transition from the resonant to  anti-resonant behavior can be noticed;
the other parameters: $\tau =0.2, J=-1$.}
\end{figure}

The space of parameters is covered by $\tau$ (ring-dot coupling),
$J$ (exchange) and $E_F$ (the Fermi level imposed by the external leads).
The multitude of parameters may give rise to different regimes
and as an illustration we show in Fig.2 ~a map of the nonflip-conductance
for two different values of $E_F$.
One notices  the change from a resonant behavior (panel a) to an
anti-resonant one (panel b). At the same time, we have found that the
flip-conductance keeps the resonant character for both values of the
Fermi energy.

The understanding of the 3D picture in Fig.2 is relatively simple;
one has to analyse separately the role played by  the ring and by the dot in
the transmittance process.
Let us call $T_{\sigma\sigma'}^{(I)}$ the transmittance of the device 
in the case $\tau=0$
(when the dot is completely detached from the ring) and express 
formally the transmittance as the sum of two contributions :
\begin{equation}
T_{\sigma\sigma'}(E,\Phi)= T_{\sigma\sigma'}^{(I)} + {\tau}^2~
T_{\sigma\sigma'}^{(II)},
\end{equation}
where the second term is added when $\tau\ne 0$. Obviously, 
$T_{\sigma\sigma'}^{(I)}$ does not depend on the flux but
may depend on the Fermi energy. Let us consider two situations :

a) the energy  of the incident electron is such that 
$T_{\uparrow\uparrow}^{(I)}\approx 0$;
then the second term describes the  resonances of the ring-dot system,
with a gate potential applied on the dot only.

b) the energy  is such that  $T_{\uparrow\uparrow}^{(I)}\lesssim 1$; 
since the total transmittance cannot be larger than $1$, 
the only possible effect which can be produced by 
$T_{\uparrow\uparrow}^{(II)}$ is an antiresonance.

We succeeded to catch these two extreme cases in Fig.2a and b, respectively,
by choosing two different values of the Fermi energy. 
More complicated 3D patterns appear in intermediate situations.
In what concerns the flip transmittance, obviously 
$T_{\uparrow\downarrow}^{(I)}=0$  so that only the situation in
panel (a) may occur and the flip transmittance will show always a 
resonant character.
\section{III. Singlet-triplet representation for the flip and non-flip 
propagators.}

In what follows we shall explain the Singlet-Triplet structure of the 
energy spectrum, the implication of the spectrum for the transport 
properties and the physical conditions ($T=0$ and 'empty band' 
ground state \cite{Nolting}) under which one can get rigorous 
results.

For S=1/2 the natural formulation of the problem is in terms of
singlet-triplet operators (see for instance \cite{Ho}).
 One can define the  fermion operators $\{d^{\dagger}_\uparrow, 
d^{\dagger}_\downarrow, d_\uparrow,d_\downarrow\}$
for the localized spin, provided one projects out all the
states with occupancy different from $1$, i.e. one has
\begin{eqnarray}
n_d=d^{\dagger}_\uparrow d_\uparrow +d^{\dagger}_\downarrow d_\downarrow =1.
\end{eqnarray}

Then, $S^z_n,~ S^+_n and~ S^-_n$ in eq (2) can be written as:
\begin{eqnarray}
S^z_n={1\over2}(d^{\dagger}_\uparrow d_\uparrow -d^{\dagger}_\downarrow
d_\downarrow)~~~~~~\nonumber\\
S^+_n=d^{\dagger}_{\uparrow} d_{\downarrow},~~ S^-_n=d^{\dagger}_{\downarrow} 
d_{\uparrow}~.
\end{eqnarray}
Simultaneously we introduce the singlet operator $\Sigma_i$ and the triplet
operators $T_i^p$ (p=1,2,3)
\begin{eqnarray}
\Sigma_i={1\over\sqrt2}~(d_{\uparrow} c_{i{\downarrow}}-d_{\downarrow}
c_{i{\uparrow}})\nonumber\\
T_i^1={1\over\sqrt2}~(d_{\uparrow} c_{i{\downarrow}}+d_{\downarrow}
c_{i{\uparrow}})\nonumber\\
T_i^2=d_{\uparrow} c_{i{\uparrow}}~~~~~~~~~~~~~~~~~~\nonumber\\
T_i^3=d_{\downarrow} c_{i{\downarrow}}~.~~~~~~~~~~~~~~~~
\end{eqnarray}

The above operators may be used in order to write the Hamiltonian (2)
in the following way (where we have used also 
$n_d=d^{\dagger}_\uparrow d_\uparrow +
d^{\dagger}_\downarrow d_\downarrow =1$):
\begin{eqnarray}
H= \sum_{\langle i,j\rangle}~(w^{\Sigma}_{ij}~\Sigma_{i}^\dagger \Sigma_{j}+w^T_{
ij}~\sum_{p}~T^{p\dagger}_{i}T^p_j)\nonumber\\
w^{\Sigma}_{ij}=w_{ij}~ e^{i2\pi\Phi_{ij}}+\delta_{ij}Vg+{3\over2 }J \delta_{ij}
\delta_{in}\nonumber\\
w^T_{ij}=w_{ij}~ e^{i2\pi\Phi_{ij}}+\delta_{ij}Vg-{1\over2}J \delta_{ij}
\delta_{in}~.
\end{eqnarray}

In the representation of $\{c,d\}$-fermions, the propagation of the free 
electron corresponding to non-flip processes
a) and b) and to the spin-flip process c) is described by the following
zero temperature two-time Green functions:
\begin{eqnarray}
G^a_{ij}(E)=\ll d_{\uparrow}c_{i{\uparrow}};c^{\dagger}_{j{\uparrow}}d^{\dagger}
_{\uparrow}\gg_{E}\nonumber \\
G^b_{ij}(E)=\ll d_{\downarrow}c_{i{\uparrow}};c^{\dagger}_{j{\uparrow}}d^{\dagger}_{\downarrow}\gg_{E} \\
G^c_{ij}(E)=\ll d_{\uparrow}c_{i{\downarrow}};c^{\dagger}_{j{\uparrow}}d^{\dagger}_{\downarrow}\gg_{E},\nonumber
\end{eqnarray}

where the ground state 
on which the Green functions are defined should be explained.
From now on the state $|>_{0}$, on which the Green functions are defined, will 
be  chosen as the vacuum with respect to both c- and d- fermions.
This is the so-called 'empty band' approach for the electronic problem. 
With this choice the propagators (16) describe the propagation of a
 "test" electron through the 
interferometer containing the localized spin.
 The same single electron approach is considered by  Menezes et al 
\cite{Menezes} and Joshi et al \cite{Joshi} when using the wave
guide approach for the ring+flipper problem in the continuum representation
(they do not emphasise the natural singlet-triplet formulation of the 
problem). Using the same approach Nolting obtained rigorous results for the
the electron excitation spectrum of ferromagnetic semiconductors \cite{Nolting}.

The generic equation of motion satisfied by these functions is 
(see the discussion in Appendix):
\begin{equation}
E\ll A;B \gg_{E}=<[A,B]_{+}>_{0}+\ll[A,H^{eff}]_{-};B\gg_{E}. 
\end{equation}
As an example, the equation of motion for $G^c$ reads:
\begin{eqnarray}
(E-V_g)G^c_{ij}=\sum_{k} w_{ik}(\Phi)G^c_{kj}+\tau_0^2 e^{-iq}\sum_{\alpha}
\delta_{i\alpha}G^c_{ij}\nonumber~~\\
+\delta_{in}{J\over2}G^c_{ij} -\delta_{in}JG^b_{ij}
-\delta_{in}{J\over2} \ll d^{\dagger}_{\downarrow}d_{\downarrow}d_{\uparrow} 
c_{i{\downarrow}};c^{\dagger}_{j{\uparrow}}d^{\dagger}_{\downarrow}\gg\nonumber\\
+J\ll c^{\dagger}_{n{\downarrow}}c_{n{\uparrow}}c_{i{\downarrow}}d_\downarrow;
c^{\dagger}_{j\uparrow}d^{\dagger}_{\downarrow}\gg
+J\ll d^{\dagger}_{\uparrow}d_{\uparrow}d_{\downarrow}c_{i\uparrow};
c^{\dagger}_{j{\uparrow}}d^{\dagger}_{\downarrow}\gg \nonumber\\
-{J\over2} \ll d_{\uparrow}(~c^{\dagger}_{n{\uparrow}}c_{n{\uparrow}} -
c^{\dagger}_{n{\downarrow}}c_{n{\downarrow}}) c_{i{\downarrow}}; c^{\dagger}_{j
\uparrow}d^{\dagger}_{\downarrow}\gg ~~~~~~~~~~~~~~~~
\end{eqnarray}
With our  choice for the ground stare, exact results are obtained as the 
equation of motion get closed without any approximation.
 Indeed, now all "higher" Green functions 
\begin{eqnarray}
\ll d_{\uparrow}(~c^{\dagger}_{n{\uparrow}}c_{n{\uparrow}} -
c^{\dagger}_{n{\downarrow}}c_{n{\downarrow}}) c_{i{\downarrow}}; 
c^{\dagger}_{j\uparrow}d^{\dagger}_{\downarrow}\gg,\nonumber\\
 \ll c^{\dagger}_{n{\downarrow}}c_{n{\uparrow}}c_{i{\downarrow}}
d_\downarrow;c^{\dagger}_{j\uparrow}d^{\dagger}_{\downarrow}\gg , 
\end{eqnarray}
etc, vanish.

We remind that if the Fermi sea is considered as the ground state, the
"higher" Green functions do not vanish any more and the solution can 
be found only approximately, using specific decoupling procedures 
( see, for instance \cite{Nagaoka,Ho}).

Eq.(18) becomes :
\begin{eqnarray}
\sum_{k}\big[\big(E-V_g-\tau_{0}^2 e^{-iq}\sum_{\alpha}\delta_{i\alpha}
-{J\over2} \delta_{in}\big)\delta_{ik}-w_{ik}\big]G^c_{kj}(E)\nonumber\\
=-\delta_{in}J~G^b_{ij}(E)~ .~~~~~~~~
\end{eqnarray}
One can verify that ${<\Sigma_i~ H~ T_j^{p\dagger}>}_{0}=0$,
meaning that the
Hilbert space $\mathcal H^{\sigma}_{2N}~\otimes~\mathcal H^S_2$
 splits as follows:
 $$\mathcal H^{\sigma}_{2N}~\otimes~\mathcal H^S_2= \mathcal H^{\Sigma}_{N}
\oplus\mathcal H^{T}_{3N}. $$
The singlet and triplet-type eigenenergies can be obtained by the
diagonalization of the matrices $w^{\Sigma}_{ij}$ and $w^{T}_{ij}$,
respectively. 

It is useful to define the singlet and triplet propagators
$$ G^{\Sigma}_{ij}(E)= \ll \Sigma_i; \Sigma^{\dagger}_j\gg_E,~
 G^{T^{(p)}}_{ij}(E)= \ll T^{(p)}_i; T^{(p)\dagger}_j \gg_E$$
as these functions satisfy   decoupled equations of motion:
\begin{eqnarray}
\sum_{k=1}^N \big[(E \delta_{ik}-w^{\Sigma}_{ik}-\tau_{0}^2 e^{-iq}
\sum_{\alpha}\delta_{i\alpha}\delta_{ik}
\big] G^{\Sigma}_{kj}(E)= \delta_{ij},\\
\sum_{k=1}^N \big[E \delta_{ik}-w^{T}_{ik}-\tau_{0}^2 e^{-iq}
\sum_{\alpha}\delta_{i\alpha}\delta_{ik}
\big] G^{T^{(p)}}_{kj}(E)= \delta_{ij},
\end{eqnarray}
where $w^{\Sigma}$ and $w^{T}$ are defined in eq(17).
The Green functions for the three triplet states are identical,
so we shall skip the upper index of $T$.
When obtaining  equations (21) and (22) the symmetry relations
\begin{eqnarray}
\ll d_{\downarrow} c_{i\uparrow}; c^{\dagger}_{j\downarrow}d^{\dagger}
_{\uparrow}\gg =
\ll d_{\uparrow} c_{i\downarrow}; c^{\dagger}_{j\uparrow}d^{\dagger}
_{\downarrow}\gg \nonumber\\
\ll d_{\downarrow} c_{i\uparrow}; c^{\dagger}_{j\uparrow}d^{\dagger}
_{\downarrow}\gg =
\ll d_{\uparrow} c_{i\downarrow}; c^{\dagger}_{j\downarrow}d^{\dagger}
_{\uparrow}\gg \nonumber
\end{eqnarray}
have been used.

It turns out from their  definitions  that the three
propagators describing the processes a), b) and c) can be expressed in terms
of singlet and triplet propagators as follows
\begin{eqnarray}
G^a= G^T \nonumber~~~~~~~~~~~~~~~\\
G^b= (G^T+G^{\Sigma})/2~~ \\
G^c= (G^T-G^{\Sigma})/2~~ \nonumber
\end{eqnarray}
The above relations show that the non-flip process a) is  purely triplet-type
since only the matrix $w^{T}$ enters the dynamics of the propagator $G^{a}$,
while the non-flip process b) and the flip process c) 
involve
both the triplet and singlet states.
We  note that from (22) we recover the  expression of $G^T$ given by (10)
in the previous section. 
\section {IV. The simplest spin interferometer}

By the use of eq.(21-22) and (23) it is now very easy to analyse 
analytically the simple example of a triangular spin interferometer 
placed in magnetic field. In this way we  get hints for understanding
the complex system.

We present first the energy spectrum (with Singlet-Triplet structure)
versus the enclosed flux (Fig.3).
The distance between a Singlet-Triplet pair of levels is, 
of course, controlled by
the exchange parameter , but there are also
degeneracy points which are independent of $J$ as we shall explain below. 
\begin{figure}
\vspace*{-17.2cm}
\epsfxsize 12 cm
\epsffile{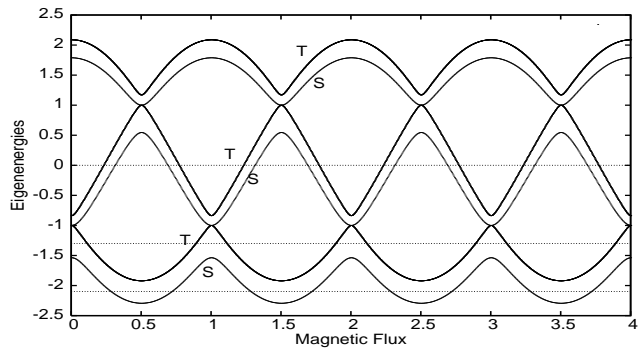}
\vspace*{4.5cm}
\caption
{The energy spectrum of the simplest (triangular) spin interferometer as
function of the magnetic flux. The singlet (S) and the triplet (T) eigenvalues
are indicated.  The three horizontal lines represent possible positions 
of $E_f$; exchange parameter $J=-0.5$.}
\end{figure}

In order to study the transport, we connect ideal leads to the sites we 
shall refer to as "1" and "3" using the hopping constant $\tau_0$. 
The flipper is located on the site "2". The sites of the triangle are 
interconnected by the hopping parameter $e^{\pm i 2\pi \Phi/3}$ 
(the total encircled flux  $\Phi$ is measured in quantum flux units).
We find the solution of eq.(21-22) as :
\begin{equation}
G_{13}^{\Sigma,T}(E)=e^{i2\pi\Phi/3}~~~
\frac{ e^{-i2\pi\Phi}+(E-\alpha^{\Sigma,T})}{\Delta},
\end{equation}
\begin{eqnarray}
\Delta=(E-\tau_0^2e^{-iq})^2(E-\alpha^{\Sigma,T})-2(E-\tau_0^2e^{-iq}) 
\nonumber\\
-(E-\alpha^{\Sigma,T})-2cos(2\pi\Phi),~~~~~~~~~~~~~~~~~~~~~~
\end{eqnarray}
with $\alpha^\Sigma=J\frac32$ and $\alpha^T=-J\frac12$ .
One can notice that, while the module of $G^{\Sigma,T}(E)$ is 
symmetrical under flux reversal and has the $\Phi_0=1$ flux periodicity, 
its phase is {\it not} symmetrical under flux reversal and has a $3\Phi_0$ 
periodicity due to the factor $e^{i2\pi \Phi /3}$.
On the other hand, the phase difference between $G^T$ and $G^{\Sigma}$
\big(i.e., $\Delta\lambda_{T\Sigma}=
atan(Im G^T/Re G^T) -
atan (Im G^{\Sigma}/Re G^{\Sigma})
$\big), 
which is important for the processes b) and c) (as expressed by eq.23), is 
periodic with  $\Phi_0$ but remains asymmetric under flux reversal. 
This explains why the transmittances assigned to the processes b) and c) are 
 asymmetric under flux reversal, as also found by Joshi et al\cite{Joshi}.

For $\tau_0=0$ the zeros of $\Delta$ in Eq.27 give the energy spectrum.
For integer (semi-integer) values of the flux the energy $E=-1 (E=1)$ 
is an eigenvalue  independent of $\alpha^{\Sigma,T}$ meaning that 
the triplet and the singlet are degenerated as it can be seen in Fig.3.

Looking at  the spectrum, we can realize that $G^{\Sigma}$ and  $G^{T}$ 
(that is also the propagator for the a) process ) will show, each of them, 
one or two peaks per period, depending on the $E_f$ (see 
the horizontal lines in the spectrum figure). 
As a result, $G^c$ and $G^b$ may show up to four peaks per period
 in the case of the week coupling with the leads.
However, for strong coupling, neighbouring singlet and 
triplet peaks will overlap with either constructive or destructive effect, 
depending on the phase difference $\Delta\lambda_{T\Sigma}$ (see Fig.4c).

As the magnetic flux is changed, the T and S levels cross the Fermi 
energy giving rise to the addition or loss of one electron at each cross. 
Taking into account that the addition of an electron corresponds to a 
phase increase of $\pi$ 
(and vice versa) the evolution of $\Delta\lambda_{T\Sigma}$ in Fig.4a is
easily understood, when the peaks are well separated in the case of weak
coupling. As we increase the coupling, the peaks are broadening and 
we can see in Fig.4c that the first pair of peaks evolves into a single 
big peak while the second pair vanishes corresponding to a phase difference
$\Delta\lambda_{T\Sigma}\approx 0$.

\section{V. The spectrum and spin transmittance of the ring+2D dot 
interferometer}
In the following we shall continue the numerical analysis of the
real system consisting of the ring+2D quantum dot (with magnetic impurity).
In order to show the complexity of the  problem,
we present first a detail of the energy spectrum of our system (Fig.5).  
In Fig.5a the dot is decoupled from the ring; the levels of the 2D finite dot 
(shown in solid line) depend on the magnetic field and can be indexed
as Triplet and Singlet levels, while 
the levels of the truncated ring (dotted line) are invariant with $\Phi$. 
\begin{figure}
\vspace*{-1.0cm}
\hspace*{-0.8cm}
\epsfxsize 12 cm
\epsffile{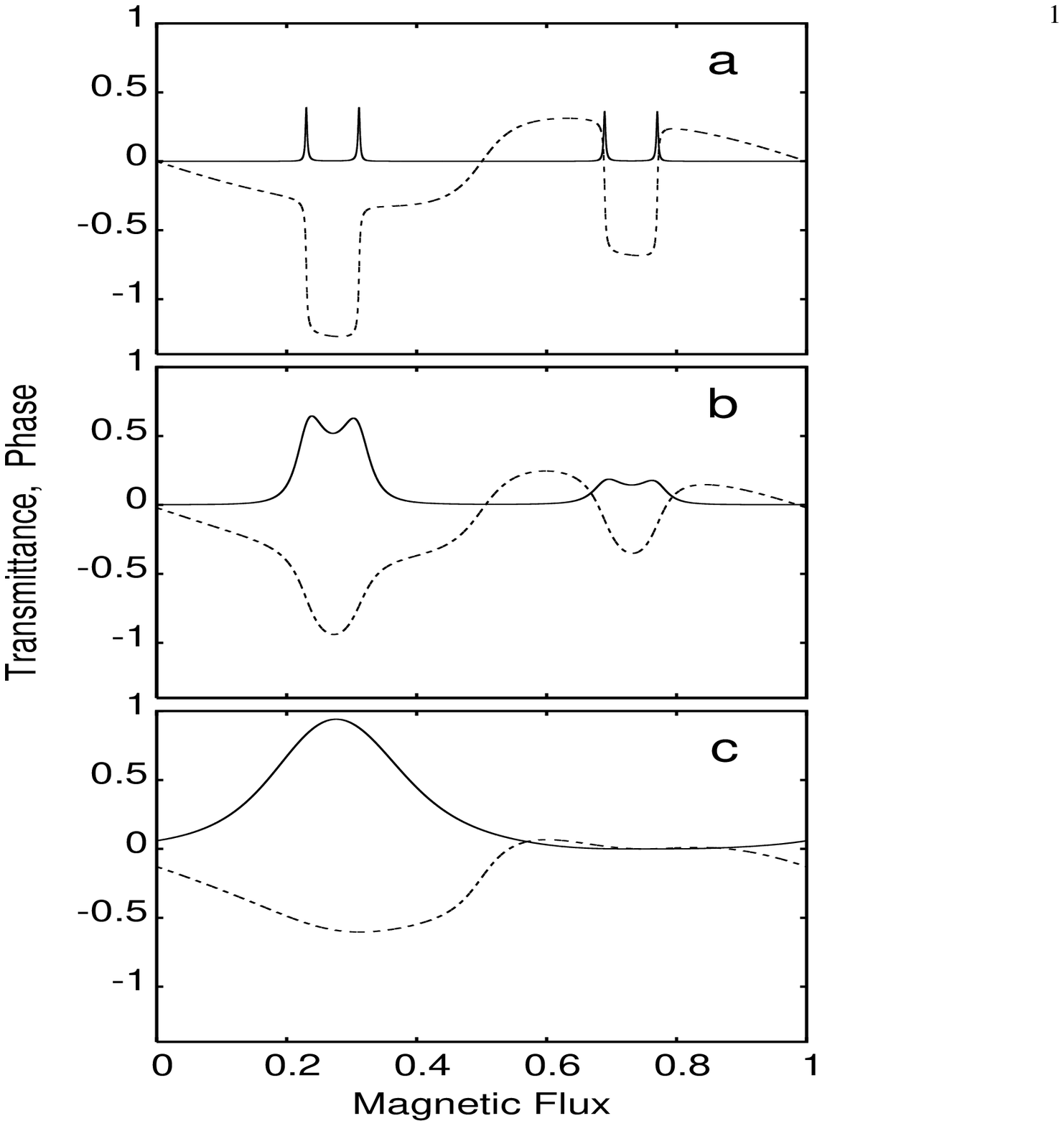}
\vspace*{-6.0cm}
\caption
{Flip transmittance (solid line) for the triangular spin interferometer, 
plotted together with the phase difference $\Delta\lambda_{T\Sigma}$
(dashed line, in $\pi$ units) for three couplings with the leads : 
a)$\tau_0=0.2$, b)$\tau_0=0.4$, c)$\tau_0=1.0$;
the other parameters are $J=-0.5$, $E_f=0$.}
\end{figure}
Obviously, when we increase the 
ring-dot coupling $\tau$ the levels undergo hybridization. 
As long as the ring-dot hybridization is  weak (as in Fig.5b)
and we are sufficiently far
away from the self-avoiding  points, the eigenvalues can still be identified as
ring-like or dot-like.
The ring-like levels begin to oscillate with
the (approximate) $\Phi_0$ period and each one splits also into a 
Singlet and a Triplet; obviously the splitting depends on the value of
the exchange $J$, and on the hybridization determined by the
 ring-dot coupling $\tau$ and on the relative distance between levels.

For strong hybridization (see Fig.5c)
the dot-like levels
begin also  to exhibit  $\Phi_0$ oscillations superimposed over
their usual slow dependence on the flux observed in Fig.5a. 
One can notice that the dot-like levels are intercalated between 
singlet-triplet pairs of ring-like levels; this  occurs because 
the exchange splitting of the dot-like levels is much bigger than 
the splitting of the ring-like levels.
In fact, for strong ring-dot coupling, the spectrum becomes complicated 
and it is difficult to identify any more the origin of different eigenvalues.
\begin{figure}
\vspace*{-1.0cm}
\hspace*{-1.0cm}
\epsfxsize 12 cm
\epsffile{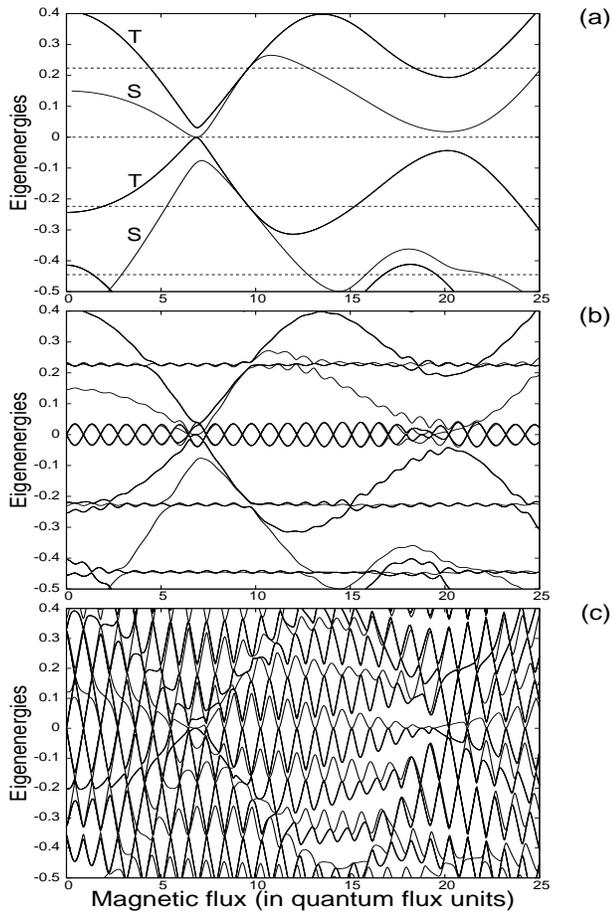}
\vspace*{-4.0cm}
\caption
{Spectrum of the ring-dot (with magnetic impurity) system for the cases:
a) The dot is decoupled from the ring ($\tau=0$). The levels of the 2D dot  
(solid line) show  flux dependence and  strong Singlet(S)-Triplet(T) splitting
; the levels of the truncated ring (dotted line) are constant.
b) Moderate ring-dot coupling ($\tau=0.4$) . Both dot and ring levels 
undergo hybridization.
The levels of the ring have a (approximate) $\Phi_0$ oscillation period and 
a very small S-T splitting that becomes more visible in the case
c) Strong ring-dot coupling ($\tau=1.0$) with strong hybridization of levels.
}
\end{figure}

Now we continue the numerical analysis of the spin transmittance in the limit
of perfect ring-leads coupling ($\tau_0=1$), which meets the usual
experimental conditions for studying the AB oscillations. 

Since the spin-up and the spin-down waves cannot interfere, being orthogonal
and corresponding to two different channels, 
the  flip-interference process can  only take place between the
flipped electron wave transmitted through the dot and the flipped wave 
reflected by the dot; this introduces a difference in phase
compared to the nonflip process and explains why the
flip-interference may be destructive at values of the magnetic field where
the direct (nonflip)process is constructive.
Another effect is the accumulation of phase in the dot which will affect the
Fourier spectra of the spin-dependent magnetoconductance fluctuations.

The interference pattern as function of the magnetic flux through the
ring depends specifically on the characteristics of the quantum dot,
namely, by its coupling to the ring and  the properties of the energy spectrum
described above . 
Implicitly, the position of the Fermi level becomes important.
\begin{figure}
\vspace*{-14.5cm}
\epsfxsize 10 cm
\epsffile{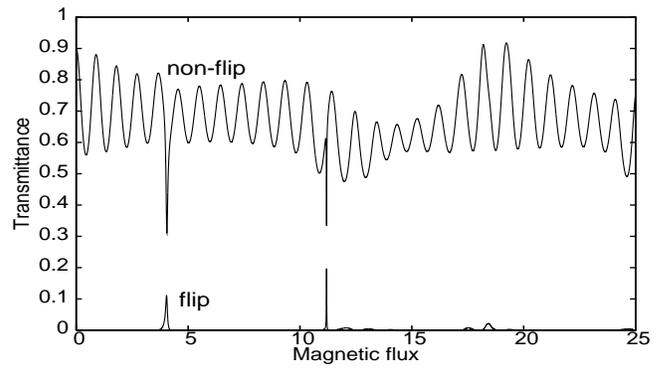}
\vspace*{5.0cm}
\caption
{The two components -nonflip and flip- of the transmittance as function of the
magnetic flux $\Phi$ through the ring
(parameters: $\tau=0.4, J=-1, V_{gate}=0, E_{F}=-0.35$).
The non-flip transmittance exhibits usual oscillations, while the flip
has peaks when the dot levels cross the Fermi energy.}
\end{figure}

For weak ring-dot coupling, 
the AB oscillations come from the participation in transport of the
ring-like levels that have a (approximate) $\Phi_0$ periodicity and  show
a Singlet-Triplet splitting (the case shown in Fig.5b). 
This splitting  is however very small since the
localized magnetic impurity is placed in the dot and acts like a small
perturbation for the "ring" levels. Then, one can
assume that $|G^T|\approx |G^{\Sigma}|$ and  the phase difference
$\lambda_{T,\Sigma} \approx 0$, except for those intervals of flux
where the dot levels get involved in transport. 
As a result the flip oscillations,
being proportional to $|G^{T}-G^{\Sigma}|$, are much smaller than the
nonflip oscillations given by $G^a$ or $G^b$. 

The position of the Fermi energy is very important for the general aspect of
the oscillation pattern. When the Fermi energy crosses (or is close to) the
oscillating ring-like levels, both the nonflip and flip transmittances show
regular oscillations (much reduced for the flip case), 
with anomalies in  those domains of the flux where 
dot-like levels are also involved in transport .

Even if the Fermi energy is far from the ring levels the nonflip 
transmittance behaves similarly.
On the contrary, the flip transmittance does not show oscillations but
has only peaks at values of the flux where the dot levels cross $E_F$. 
In this way the energy levels of the dot can be identified by the
flip transmittance. We illustrate this situation in Fig.6.
\begin{figure}
\vspace*{-2.0cm}
\hspace*{-1.0cm}
\epsfxsize 10 cm
\epsffile{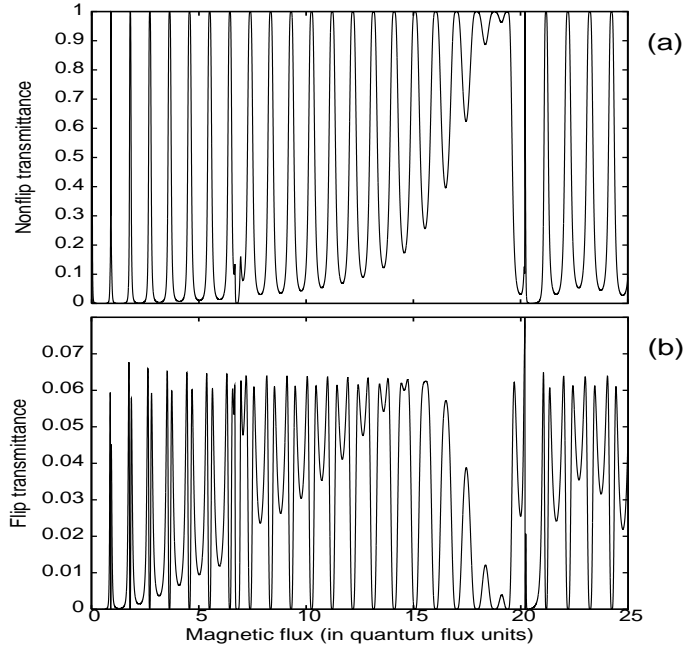}
\vspace*{-3.2cm}
\caption
{The two components of the transmittance as function of the
magnetic flux $\Phi$ through the ring: (a) nonflip  
$T_{\uparrow\uparrow}$, (b) flip $T_{\uparrow\downarrow}$ 
(parameters: $\tau=1, J=-1, V_{gate}=0, E_{F}=0$). Note the double peak
structure of the flip and the complementary windows in the range
[15-20$\Phi_0$].}
\end{figure}

For the strong ring-dot coupling $\tau=1$, the spectrum shown in Fig.5c 
plays the game and
the result consists in strong oscillations for any Fermi level. Four peaks
per period cannot be observed since the splitting is small compared to the
broadening induced by the strong coupling of the system to the leads $
\tau_0=1$.
The general picture is that one of single or double peak per period
both for the flip and non-flip transmittance, depending however on the 
$E_F$. The double peak is usually asymmetric as in Fig.4b of the
triangle model.

A peculiar situation, 
when the spin effect is strong, 
occurs at $E_F=0$ and is shown in Fig.7 .
	The difference noticed between the panels (I) and (II)
is due to the fact that the scattering 
processes (b=nonflip) and (c=flip) combine in different 
ways the singlet and triplet
propagators. The situation is presented at a better resolution in Fig.8a,
and it can be noticed that $T_{\uparrow\downarrow}$ is nearly zero in
the same place where $T_{\uparrow\uparrow}\approx 1$. With necessity,
the quasi-zero of the flip transmittance occurs when $|G^T|\approx 
|G^{\Sigma}|$ and $\lambda_{T,\Sigma}\approx 0$. Under the same circumstances,
the nonflip process (b) described by $|G^T+G^{\Sigma}|$ has a maximum value.
This explains the destructive interference of he flip process at the same
value of the flux where non-flip interference process is constructive.
It can also be noticed that 
the transmittance varies strongly in some domains
(as for instance between $15-20\Phi_0$ ). 
A separate calculation indicates that the transmission of
the dot is much suppressed in these domains,
while the amplitude of $T_{\uparrow\uparrow}$ through the lower arm is
close to 1; these two facts yield the picture shown in Fig.7a.
On the other hand, when the transmission through the dot is low, 
the probability of the flip process is reduced explaining the complementary 
window of the $T_{\uparrow\downarrow}$ in Fig.7b.

We have checked that such windows are a combined effect of the 
presence of the dot and of the magnetic impurity; it turns out, however that
for $E_F=0$ the windows in the interference pattern comes from the 
exchange. 

One can observe in the  top panel of the Fig.8 that the large peak
representing the  nonflip-interference embraces the double peak 
representing the flip-process. 
The splitting increases with the exchange $J$  and hence the width of the
large peak depends also on this parameter.

An important instrument for the analysis of the transmittance oscillations
is the Fourier transform \cite{WW}.
The Fourier spectra of  $T_{\uparrow\uparrow}$ and $T_{\uparrow \downarrow}$
are shown in Fig.8b and  Fig.8c in the case of the perfect coupling between 
the dot and the ring ($\tau=1$).
The fact that already the first interference process occurs with the
participation of a reflected wave results in the reduction of the first 
maximum in the Fourier spectrum for the flip-channel.

The coupling $\tau$ is important for the aspect of the Fourier spectrum
because any constriction at the dot gives rise
to a reflection of the spin-up electron, and changes the aspect
of the Fourier spectrum of $T_{\uparrow\uparrow}$,
that might become  similar to the spectrum of $T_{\uparrow\downarrow}$.

When the dot is attached to the ring the positions of the interference 
peaks are shifted due to an accumulation of phase in the dot; this
effect is clearly visible in the Fourier spectrum of the AB oscillations,
namely the positions of the maxima do not 
coincide any more with the integer multiples of the $2\pi/\Phi_0$,
which are characteristic to the individual ring. The shift can be noticed
both in Fig.8b and Fig.8c for the non-flip and flip transmittance, respectively.
\begin{figure}
\vspace*{-1.6cm}
\hspace*{-0.5cm}
\epsfxsize 13 cm
\epsffile{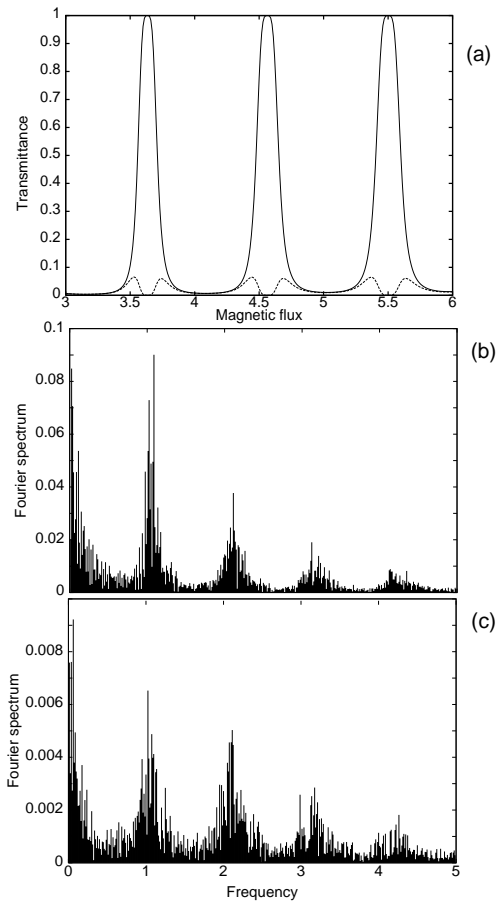}
\vspace*{-3.6cm}
\caption
{(a) The double-peak structure of $T_{\uparrow\downarrow}$ embraced by 
the large peak of  $T_{\uparrow\uparrow}$ .
The Fourier spectrum corresponding to the curves shown in FIG.3:
(b) the nonflip component. (c)
the flip component shows more noise and a reduced value of the 
first maximum as the result of the specific  interference process
(parameters: $\tau=1, J=-1,V_{gate}=0, E_{F}=0$).}
\end{figure}
In the mesoscopic physics, the Fano effect is the consequence
of the hybridization between the continuum spectrum of the
leads and the discrete spectrum of the meso-system \cite{Xu}.
The asymmetric shape of the transmittance peaks, the occurrence of zeros,
and the behavior of the transmittance phase on the resonances and between them 
are the aspects of interest.

In the system under consideration, the dependence on the gate potential 
(at fixed magnetic field) of the transmittance phases corresponding 
to different flip and non-flip processes 
show the usual behavior and we shall not insist on these aspects:
the phase increases with $\pi$ on each resonance and there are also 
discontinuous jumps 
with $\pm\pi$ between the resonances with the same parity,
where the transmittance  vanishes \cite{YB}.
Since in our case the magnetic field is non-zero, these jumps 
do not represent  true Fano zeros because
the real and imaginary part of the transmittance amplitude cannot
change simultaneously the sign. Nevertheless, when the change in sign 
of $Re T$ and $Im T$ occurs at close values of the gate bias,
the variation of the phase is still abrupt. Then,
the jump of the phase can be upwards (with a gain of $\pi$) 
or downwards (with a loss of  $\pi$); obviously, the difference is given
by the sign of $Im T$ at $ReT =0$ (which is positive in the first case 
and negative in the second one). 

On the other hand, the phase have a very complex behavior as function
of magnetic field.
Some analysis can be made for the simple triangle model, where we can
study separately the phase of the denominator and of the numerator in
Eq.(24).
The two contributions are easy to be identified for $\tau_0<<1$ when the
resonances are narrow. Then the phase of the denominator $\Delta$ (eq(25)) 
has a sudden evolution with $\pi$ and $-\pi$ on consecutive resonances, 
but remains constant in between.
On the other hand, the phase of the numerator, which in fact does not
depend on $\tau_0$, has a very small evolution on the narrow resonances
and is responsible for the monotonic increase (or decrease, depending on
$E_F$) of the phase  of $G_{13}^{\Sigma,T}$ between
resonances. We note that the numerator of these propagators
vanishes only incidentally for $E-\alpha^{\Sigma,T}=\mp1$ and integer 
(semi-integer) values of the flux,
In the limit $\tau_0\rightarrow 1$  the width of the resonances increases
causing the overlap of neighbouring resonances and a 'bump' instead of
a $\mp\pi$ evolution of the phase at the variation of the flux.

The non-monotonic variation of the phase on the peak indicates that
the phasor of the transmittance in the complex plane ($ReT, ImT$) has a
turning point (as shown in Fig.9b and Fig.9d).

When extending this analysis to the complex system we realize that 
the numerator becomes a very complicated trigonometric function and 
exact zeros, and even quasi-zeros, may exist only incidentally. 
So, the general aspect of the phase consists in a smooth dependence 
on flux between two consecutive peaks and a bump on each peak. 

The special situation when the phase indicates quasi Fano zeros has been found 
numerically at $E_F=0$ for the case of impurity free dot ($J=0$); 
 however, the usual behavior is recovered in the presence of the exchange 
($J\ne 0$).
The two situations are shown in Fig.9. In the same figure we give also the 
phase of the process (c), which is obtained by combining the phases
and modules of the $G^T$ and $G^{\Sigma}$ for the complex ring-dot system.
\begin{figure}
\vspace*{-1.1cm}
\hspace*{-0.5cm}
\epsfxsize 10 cm
\epsffile{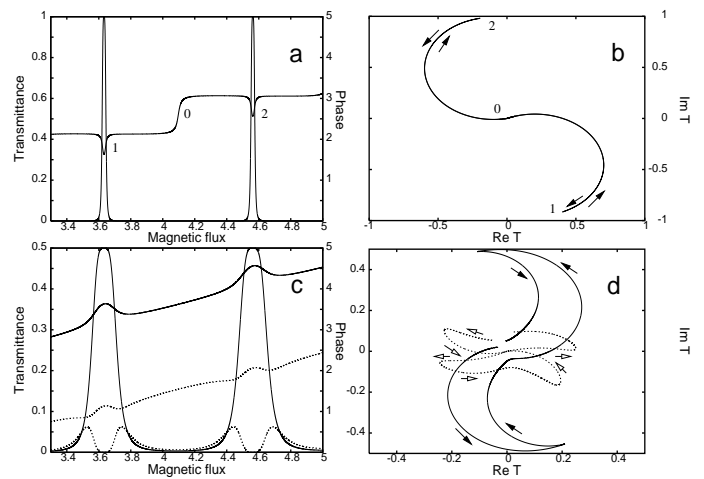}
\vspace*{-4.5cm}
\caption
{a) Transmittance and phase for the impurity-free system (J=0);
 b) The corresponding evolution of the phasor (ImT vs ReT).
The points 1,0,2 correspond to the same points in the panel (a).
The arrows indicate the evolution of the phasor with increasing flux 
in the range $[3.3-5.0\Phi_0]$.
 c) Transmittance and phase for the Triplet (solid line) and the
 Flip (dashed line); 
 d) The corresponding evolution of the phasors. The phasor 
of the Flip (amplified for better visibility)
has four turning points corresponding to the four peaks of the 
flip-transmittance in panel (c). }
\end{figure}
\section{VI. Conclusions}
In conclusion, in this paper we report the first calculation
of the spin-dependent transmittance through an interferometer consisting 
of a ring with embedded two-dimensional dot doped with magnetic impurities.
For the time being, there  are no experiments to be compared
with,  but they are expected in the near future, and our goal is to make
predictions. The study is based on the spectral analysis and a suitable
transport formalism which has been developed for the single-particle case.

The  aspect of the Aharonov-Bohm oscillations depends strongly on the Fermi
energy and the ring-dot and ring-leads coupling parameters.
Generally,  the oscillations exhibit  single or double asymmetric peaks 
per period as it is suggested by the analytically solvable triangular model.
The flip oscillations are much reduced in amplitude compared to the
nonflip ones.
The finite size of the dot, combined with the effect of the impurity,
gives rise to windows in the interference pattern where the flip and nonflip
components of the transmittance behave in a complementary way.
In the case of weak ring-dot coupling the AB oscillations of the flip
channel  may cease to exist, the flip magnetoconductance
showing instead peaks corresponding to  the levels of the dot.
The general aspect of the transmittance phase of the AB oscillations 
consists in bumps on the oscillation peaks and  a monotonic behavior 
between them.

In order to explain these properties we have expressed the different 
physical tunneling processes in terms of singlet and triplet propagators,
which is the natural formulation of the problem.

{\bf Acknowledgments}
The support of DFG /SFB 608 and  the Romanian 
CERES Programme is acknowledged. We thank M.Nita and V.Dinu for very
useful discussions.
\section {Appendix}

We introduce  the Hamiltonian of the semi-infinite non-interacting leads
$H^L$ and the lead-ring coupling $H^{LR}$ . The different leads are indexed
by $\alpha$, which stands also for the site where the lead is attached.
Then,
\begin{eqnarray}
H^L=\sum_{\alpha\sigma}\sum_{n=0}^\infty 
c_{n\sigma}^{\alpha \dagger}c_{n+1 \sigma}^{\alpha} + H.c.\\
H^{LR}=\tau_0 \sum_{\alpha \sigma}c_{0\sigma}^{\alpha\dagger}
c_{\alpha \sigma} +H.c.,
\end{eqnarray}
where $\tau_0$ is the hopping integral between the first site  on the 
lead, $n=0$, and the site $\alpha$ on the ring.

The Hamiltonian of the whole system is 
\begin{eqnarray}
H^{total}= H + H^L + H^{LR},
\end{eqnarray}
 where $H$ was defined already in (2).
Let us introduce also
\begin{eqnarray}
H^{eff}=H+\tau_0^2 e^{-iq}\sum_{\alpha \sigma}c_{\alpha \sigma}^\dagger
c_{\alpha \sigma},
\end{eqnarray}
where $E=2 cos(q)$ is the energy of the incident 'test' electron.
We have to prove  that 
\begin{eqnarray}
\ll[d_{\uparrow}c_{i\downarrow},H^{Total}];c^{\dagger}_{j{\uparrow}}
d^{\dagger}
_{\downarrow}\gg=
\ll[d_{\uparrow}c_{i\downarrow},H^{eff}];c^{\dagger}_{j{\uparrow}}d^{\dagger}
_{\downarrow}\gg \nonumber,
\end{eqnarray}
meaning that
the effective Hamiltonian $H^{eff}$ can be used instead of $H^{total}$.
This will be valid if the Green functions are defined on the 
vacuum at zero temperature.
The proof is given for 
$G^c$ but the same is true for all the functions in (16) .

Performing the commutators, one gets (we write only the non-vanishing terms):
\begin{eqnarray}
\ll\big[ d_{\uparrow}c_{i\downarrow},H^{L}+H^{LR}\big] ;
c^{\dagger}_{j{\uparrow}}d^{\dagger}_{\downarrow}\gg=
\delta_{i\alpha}\tau_0 \ll d_{\uparrow}c_{0{\downarrow}}^\alpha;
c^{\dagger}_{j{\uparrow}}d^{\dagger}_{\downarrow}\gg \nonumber 
\end{eqnarray}
and,
\begin{eqnarray}
\ll[d_{\uparrow}c_{i\downarrow},\tau_0^2 e^{-iq}\sum_{\alpha \sigma} 
c_{\alpha \sigma}^\dagger c_{\alpha \sigma}];c^{\dagger}_{j{\uparrow}}
d^{\dagger}_{\downarrow}\gg= ~~~~~~~~~~~~~~~~~~~~~~\nonumber \\
\delta_{i\alpha}\tau_0^2 e^{-iq}\ll d_{\uparrow}c_{\alpha \downarrow};
c^{\dagger}_{j{\uparrow}}d^{\dagger}_{\downarrow}\gg.\nonumber
\end{eqnarray}
In order to show that these two expressions are equal, we define
$$ g_n=: \ll d_{\uparrow}c_{n{\downarrow}}^\alpha;
c^{\dagger}_{j{\uparrow}}d^{\dagger}_{\downarrow}\gg,~~~~$$
which satisfy the following equations of motion:
\begin{eqnarray}
E g_0 = g_1+\tau_0\ll d_{\uparrow}c_{\alpha \downarrow};
c^{\dagger}_{j{\uparrow}}d^{\dagger}_{\downarrow}\gg ,~~~~~~~~~~~~\nonumber \\
E g_n = g_{n-1} + g_{n+1} ~,~~ n=1,2, ..~.~~~~~~~~~~~
\end{eqnarray}
The  solution of eqs (30) is 
\begin{equation}
g_n=\tau_0 e^{-iq(n+1)}
\ll d_{\uparrow}c_{\alpha \downarrow};
c^{\dagger}_{j{\uparrow}}d^{\dagger}_{\downarrow}\gg .~~~~~~~~~ \nonumber 
\end{equation} 
so that, finally,
\begin{equation}
\ll d_{\uparrow}c_{0{\downarrow}}^\alpha;
c^{\dagger}_{j{\uparrow}}d^{\dagger}_{\downarrow}\gg = 
\tau_0 e^{-iq}\ll d_{\uparrow}c_{\alpha\downarrow};
c^{\dagger}_{j{\uparrow}}d^{\dagger}_{\downarrow}\gg , \nonumber
\end{equation}
which concludes the demonstration that in this case the use of  $H^{eff}$
is equivalent to the use of $H^{total}$.


\begin{references}
\bibitem{comm}
for a ring of radius $50nm$, described in the TB model by 
32 sites on the ring, and for $J$ of the order of  the hopping
integral ($w\approx 5 meV$) this condition is fulfilled 
at $B<200 T$ (the
g-factor of GaAs has been considered ).
\end{references}
\end{document}